# Absorbing photonic crystals for thin film photovoltaics

O. El Daif[a,b], E. Drouard[a], G. Gomard[a,b], X. Meng[a,b], A. Kaminski[b], A. Fave[b],
M. Lemiti[b], E. Garcia Caurel[c], P. Roca i Cabarrocas[c], S. Ahn[d], H. Jeon[d], C. Seassal[a]

[a] Université de Lyon, Institut des Nanotechnologies de Lyon-INL, UMR CNRS 5270
Ecole Centrale de Lyon, F-69134 Ecully, France;
[b] Université de Lyon, Institut des Nanotechnologies de Lyon-INL, UMR CNRS 5270
INSA Lyon, F-69621 Villeurbanne, France
[c] LPICM, CNRS UMR 7647, Ecole Polytechnique, Palaiseau, France
[d] Department of Physics and Astronomy, Seoul National University, Seoul 151-747, Korea

**ABSTRACT**

The absorption of thin hydrogenated amorphous silicon layers can be efficiently enhanced through a controlled periodic patterning. Light is trapped through coupling with photonic Bloch modes of the periodic structures, which act as an absorbing planar photonic crystal. We theoretically demonstrate this absorption enhancement through one or two dimensional patterning, and show the experimental feasibility through large area holographic patterning. Numerical simulations show over 50% absorption enhancement over the part of the solar spectrum comprised between 380 and 750nm. It is experimentally confirmed by optical measurements performed on planar photonic crystals fabricated by laser holography and reactive ion etching.

## 1. INTRODUCTION

Photovoltaic (PV) devices based on thin absorbing layers constitute the so-called "second generation solar cells". With regards to the classical "first generation" devices based on thick silicon wafers, thin film solar cells offer a cheap alternative. Such approaches are based on materials which exhibit a high absorption coefficient and ease of thin layer deposition, like hydrogenated amorphous silicon (aSi:H). However, their photoelectrical conversion efficiency is limited for two main reasons: i) the reduced thickness limits sunlight absorption, and ii) the defects in the deposited materials yiekd minority photocarrier recombination, and therefore electrical losses. In such devices, increasing light trapping and absorption is therefore essential. This is all the more relevant if high absorption is achieved in a very thin layer, since the fabrication costs will then be reduced, so as bulk carrier recombination. This may be achieved using various kinds of light trapping techniques allowed by the development of nanophotonics.

Absorption enhancement using surface plasmons has been proposed during the last years, either with self-assembled nanoparticles [1] or regular arrays of nanostructures fabricated using top-down technological processes [2]. In these approaches, the absorption enhancement around the plasmon resonances is balanced by light absorption in the metallic nanostructures. Photonic Crystals (PCs) have also been considered to realize either a back reflector [3] or selective filters for tandem solar cells [4]. Thin semiconductor layers may also be patterned as planar PCs; this enables efficient light trapping and significant increase of solar light absorption over a wide spectral range [5-9]. Using this approach, we proposed designs based on the use of aSi:H PC Bloch modes standing over the light-line. The use of such slow light modes has been investigated during the past years, with a view to control light-matter interaction in the case of surface emitting lasers or other nonlinear optical devices [10].

In this paper, we discuss on the implementation of such planar PC structures in order to realize PV solar cells. The photonic structures consists of a 2D array of holes (2DPC) or a 1D lattice of air slits (1DPC) drilled in an aSi:H layer positioned between layers of moderate refractive indices materials like transparent conductive oxides (TCO). Incident light coupling and absorption of sunlight in the aSi:H layer is then controlled by the photon lifetime corresponding to the optical modes of the PC. The acceptance angle is moreover controlled by the dispersion characteristics of the PC Bloch modes. In the last section, we will demonstrate experimentally the absorption enhancement of a 95nm thick aSi:H layer deposited on glass.

## 2. SIMULATED ABSORPTION ENHANCEMENT

### 2.1 One-dimensional photonic crystal

We performed rigorous coupled wave analysis (RCWA) simulations using CAMFR [10] to obtain the absorption spectrum of a layer of aSi-:H deposited on glass, in the *380-750nm* wavelength range, as a function of the 1DPC parameters (its period *L* and its filling factor in material, *ff* defined as the ratio of the aSi:H lines width over the period).We chose to work either with *100nm* or with *400nm* aSi:H layer thickness, the thinner should indeed ensure an optimal carrier collection [15], and the thicker is closer to typical aSi:H cells. The investigation is limited to wavelengths below *750nm*, i.e. in the spectral range where aSi:H exhibits a significant absorption coefficient [14], and we considered the AM1.5G solar spectral intensity distribution. Although for clear symmetry reasons the properties of the studied structures are polarisation dependent, we considered an unpolarised light by averaging the two polarisations of an incident plane wave. The corresponding contour mapping of spectrally integrated absorption efficiencies is shown on Fig. 1. Using the same method as for the PC structure, we predicted that a simple 100nm unpatterned layer of aSi:H on glass absorbs about *30%* of the solar light intensity integrated over the same spectral range. In the case of the 1DPC, the absorption is increased to more than *42%* with *L=300nm* and *ff=68-70%*. As for the *400nm* layer thickness, the unpatterned layer absorbs 40% of this spectrum, where the planar PC absorbs more than *56%* with PC parameters of *L=400nm* and *ff=50%* or *L=350nm* and *ff=56-57%*. One can note here the integrated absorption tolerance to a change in the PC parameters: a *10%* relative change in the filling factor or in the PC period affects the integrated absorption only by *2%*.

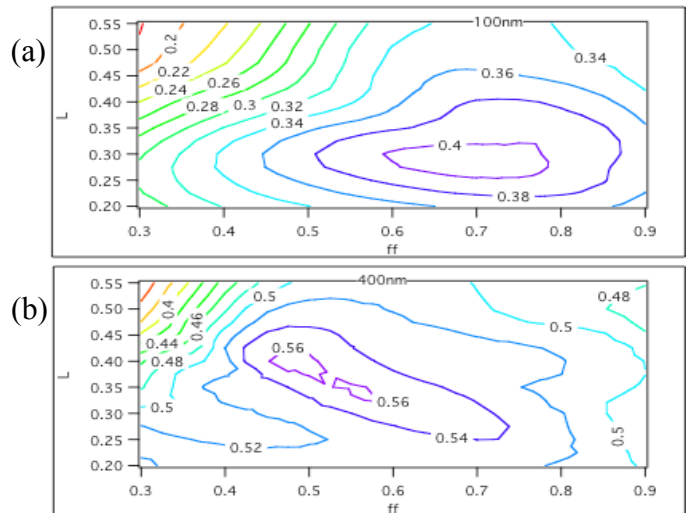

Figure 1: Integrated absorption efficiency of a 100nm thick aSi:H layer in air, as a function of the 1DPC parameters L (in micron) and ff (in %) a layer of 100nm (a) and a 400nm thick a-Si:H layer (b).

### 2.2 Two dimensional photonic crystal

We performed similar simulations for a 100nm thick 2D aSi:H planar PC. We chose to work with a square array of circular holes in order to stay within technological realism. The period is the same in the two directions of space, allowing an independence of the structure topography over the two planar spatial directions. The filling factor *ff* is now defined as the ratio of the air holes diameter over the period. The resulting integrated absorption efficiency mapping as a function of these parameters is shown on Fig. 2.

Given the high needs of these simulations in terms of resources, we limited ourselves to one single thickness and to a layer surrounded by air. One may see that a further improvement in absorption efficiency is shown, with respect to the same layer on glass but with a 1D planar PC: *55%* of the same *380-750nm* spectrum is absorbed instead of *44%* for an optimised aSi:H 1DPC also surrounded by air. This is attributed to the absence of polarisation dependence in the case of the 2D planar structure where there is an invariance of the structure with respect to the two directions of the layer plane. This plane is also the plane containing the electric and magnetic vectors of the incident plane wave.

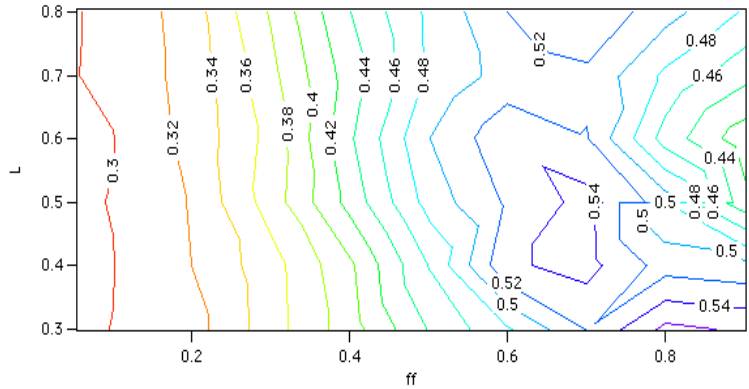

Figure 1 Integrated absorption efficiency of a 100nm thick aSi:Hlayer in air, as a function of the 2DPC parameters L (in micron) and ff (in %).

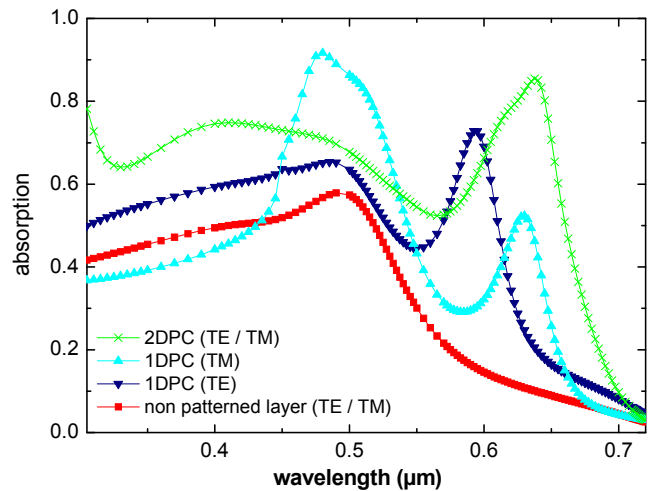

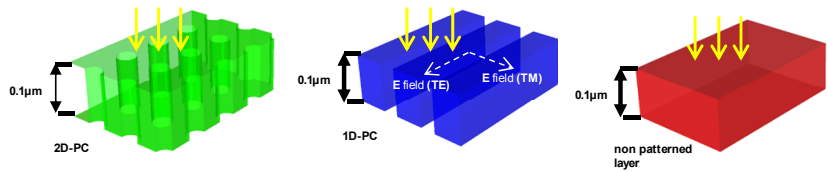

Figure 3 Effect of 1D and 2D patterning of a 100nm thick aSi:H layer in air on the absorption spectra for TE and TM polarization (optical simulation performed under normal incidence).

Figure 3 shows the absorption spectra of the various optimised structures mentioned above. The improvement in absorption thanks to patterning is clear on all parts of the spectrum despite the diminution of the quantity of material. At low wavelengths, where the material is already highly absorbing, the patterning has an antireflection coating effect. At high wavelengths, where the material's extinction coefficient $k$ decreases extremely rapidly, the effect is a resonant effect. Light is coupled in a PC Bloch mode and therefore photon lifetime in the structure is increased, increasing subsequently absorption. For 1DPC abso9rption spectra, the effect of polarisation is clearly observable and the absorption enhancement is better for TM polarisation than for TE polarisation.

## 3. DEVICE FABRICATION AND ABSORPTION MEASUREMENTS

### 3.1 Fabrication

First demonstrators consist of *95nm* thick aSi:H 1DPC structures standing on a glass substrate. aSi:H films were deposited in a standard capacitively coupled radio frequency (RF) glow discharge system [11]. In order to generate the PC pattern over a large-area at high throughput, we employed laser holography. Two coherent planar waves originated from a single 325-nm He–Cd laser were combined to interfere with each other and to produce a series of grating lines on a resist mask yielding a PC structure with a period of *335nm* and *ff =58+/-2%*. These parameters do not strictly correspond to the optimal structure, due to technological inaccuracies. The details of the holographic process can be found in previous works [12,13]. The resist mask pattern was transferred into a hard mask of silica. Amorphous silicon was then selectively etched using Reactive Ion Etching, with a gas plasma based on $SF_6$ and Ar. The silica was left above the aSi:H stripes, adding about 100nm of transparent medium, which may account for a transparent conductive oxide layer to be used for carrier collection in a complete stack.

### 3.2 Experimental results

We performed absorption measurements on the fabricated structures with an integrating sphere. A general optical absorption increase is observed over the considered spectrum, as shown on Fig.4. One may observe in particular that the absorption spectrum stays well above *10%* for wavelengths above *650nm*, while the planar structure does nearly not absorb anymore. These measurements are in agreement with the simulations shown on the same graph. They were performed on structures defined as close as possible to the experimentally measured topographical profile, although still considering square aSi:H stripes. The latter is different from the real profile, and explains the slight differences observed between theory and experiments.

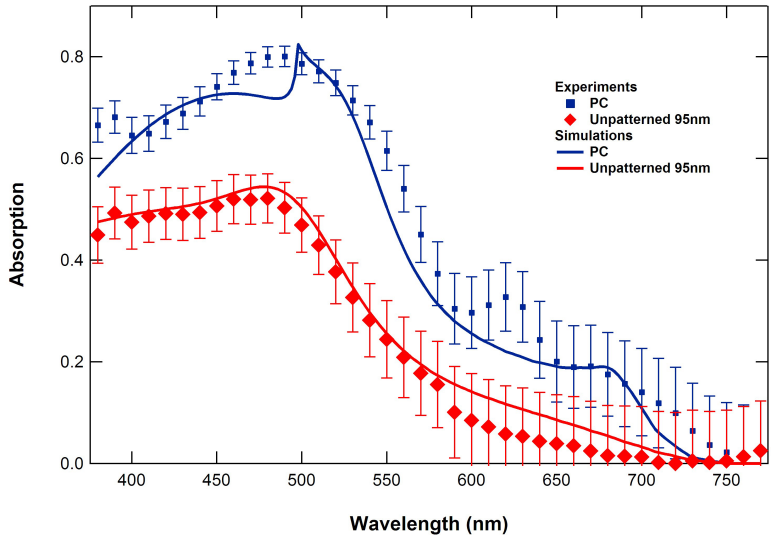

Figure 4 Measured (squares) and simulated (continuous lines) absorption spectra of a 95nm thick aSi:H layer with and without PC patterning.

## 4. CONCLUSION

Simulations made on patterned aSi:H layers revealed a significant increase of the number of absorbed photons. Compared to unpatterned layers, 1D patterning of thin amorphous silicon (aSi-:H) layers improves absorption efficiency by *50%* of the solar light over a *380-750nm* spectrum. Indeed, a layer of *100nm* of aSi:H absorbs about *30%* of this spectrum, while absorption raises to *45%* with a 1D patterning. Moreover a 2D patterning allows a *56%* of absorption efficiency, with no polarisation dependence. The experimental application of these concepts proved to be successful on a 95nm thick aSi:H layer deposited on glass. As seen through simulations, small variation of the refractive index around the absorbing layer do not change qualitatively the results, therefore integration of these absorbing PC in between two TCO layers for subsequent aSi:H thin solar cell fabrication will be beneficial for the final efficiency.

## ACKNOWLEDGEMENTS

We acknowledge funding from the French National Research Agency (ANR) Solar Photovoltaic program (SPARCS project). This work was partly performed in the frame of the French-Korean International Associated Lab "Center for Photonics and Nanostructure". Finally we thank C. Jamois, X. Letartre and P. Viktorovitch for helpful discussions, the NanoLyon Technology Platform, in particular P. Crémillieu and R. Mazurczyk for efficient support, and R. Perrin for his kind technical help.